\documentclass[twocolumn,floatfix,showpacs]{revtex4}
\usepackage{graphicx}
\usepackage{amsmath}
\begin{document}

\title{Excess conductance of a spin-filtering quantum dot}
\author{C. W. J. Beenakker}
\affiliation{Instituut-Lorentz, Universiteit Leiden, P.O. Box 9506, 2300 RA Leiden, The Netherlands}
\date{January 2006}
\begin{abstract}
The conductance $G$ of a pair of single-channel point contacts in series, one of which is a spin filter, increases from $1/2$ to $2/3\times e^2/h$ with more and more spin-flip scattering. This excess conductance was observed in a quantum dot by Zumb\"{u}hl et al., and proposed as a measure for the spin relaxation time $T_{1}$. Here we present a quantum mechanical theory for the effect in a chaotic quantum dot (mean level spacing $\Delta$, dephasing time $\tau_{\phi}$, charging energy $e^{2}/C$), in order to answer the question whether $T_{1}$ can be determined independently of $\tau_{\phi}$ and $C$. We find that this is possible in a time-reversal-symmetry-breaking magnetic field, when the average conductance follows closely the formula $\langle G\rangle=(2e^{2}/h)(T_{1}+h/\Delta)(4T_{1}+3h/\Delta)^{-1}$.
\end{abstract}
\maketitle

The study of spin relaxation in the presence of chaotic scattering is a challenge for theorists and experimentalists. The common goal is to identify transport properties that can be readily measured and that depend as directly as possible on the spin relaxation time ($T_{1}$). One line of research is to study how quantum interference effects such as weak localization or universal conductance fluctuations are modified by spin relaxation \cite{Zai05}. A direct relation with $T_{1}$ in that context is hindered by the fact that dephasing (both of the orbital and of the spin degrees of freedom) also modifies the quantum interference effects. Another line of research is to study spin-resolved current noise \cite{Sau04}. There a direct relation with $T_{1}$ is possible, but the complications involved in the measurement of both time- and  spin-dependent current fluctuations have so far prevented an experimental realization. Ideally, one would like to relate $T_{1}$ to the time averaged current in a way which is insensitive to dephasing. It is the purpose of this work to present such a relationship.

Our research was inspired by the proposal of Zumb\"{u}hl et al.\ of a new technique to measure spin relaxation times in confined systems \cite{Zum01}. These authors reported measurements of the conductance of an open two-dimensional GaAs quantum dot in a parallel magnetic field. One of the two point contacts was set to the spin-selective $e^{2}/h$ conductance plateau. The other point contact was set to transmit both spins. In this configuration, the classical series conductance of the two point contacts is $\frac{1}{2}\times e^{2}/h$ if there is no spin relaxation and $\frac{2}{3}\times e^{2}/h$ if there is strong spin relaxation. What we will show here is that the ensemble averaged conductance in a time-reversal-symmetry-breaking magnetic field varies between these two limits as a rational function of the product of $T_{1}$ and the mean level spacing $\Delta$ --- largely independent of the presence or absence of dephasing.

\begin{figure}
\centerline{\includegraphics[width=8cm]{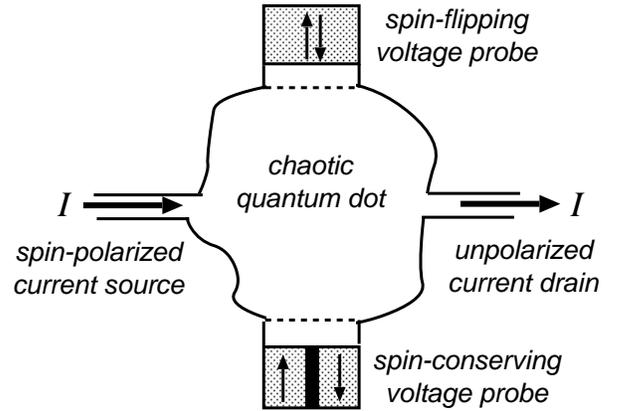}}
\caption{
Illustration of the model. A current $I$ is passed through a quantum dot via two single-channel leads, at a voltage difference $V$. Spin-flip scattering and decoherence (with relaxation times $T_{1}$ and $\tau_{\phi}$) are introduced by means of fictitious voltage probes, separated from the quantum dot by tunnel barriers (dashed lines). The lower (ferromagnetic) voltage probe reinjects an electron into the quantum dot with the same spin but a random phase (contributing only to $\tau_{\phi}$). The upper (normal metal) voltage probe randomizes both spin and phase (contributing to both $T_{1}$ and $\tau_{\phi}$).}
\label{fig_twoprobes}
\end{figure}

The geometry of the problem is sketched in Fig.\ \ref{fig_twoprobes}. We discuss its various ingredients.

Electrons in a two-dimensional electron gas (2DEG) enter and leave the quantum dot via two single-channel quantum point contacts (QPC). A QPC can operate as a spin filter in a magnetic field \cite{Sac01,Pot02}, as a result of the slightly different Fermi wave lengths of spin-up and spin-down electrons. The filtering property of a QPC can be turned on and off by adjusting its local electrostatic potential (via a gate voltage). The polarity of the spin filter is fixed by the direction of the magnetic field. The conductance becomes sensitive to spin-flip scattering if one point contact is a spin filter while the other transmits both spin directions. (To be definite, we will take the current source as the spin filter, but it does not matter which is which in the linear response regime.)

We assume that the magnetic field is sufficiently weak that we may neglect the spin dependence of the Fermi wave length inside the quantum dot (where the Fermi energy is much greater than in the point contact). The effect of the magnetic field on the orbital motion will typically break time reversal symmetry (symmetry index $\beta=2$), if the field is oriented perpendicular to the 2DEG. We contrast this with the case $\beta=1$ of preserved time reversal symmetry, appropriate for moderately weak parallel fields (until the finite thickness of the 2DEG\ drives $\beta=1\mapsto 2$ even for a parallel field \cite{Fal02}). The mean dwell time in the quantum dot is assumed to be small compared to the spin-orbit scattering time, so that spin-orbit coupling can be neglected. Landau level quantization inside the quantum dot is assumed to be insignificant. The effects of a finite charging energy will be assessed at the end of the paper.

Two independent time scales characterize the spin decay, the time scale $T_{1}$ on which the spin direction is randomized and the time scale $T_{2}\leq 2T_{1}$ on which the phase of the spin-dependent part of the wave function is randomized \cite{Eng04,Gol04,Coi05}. In closed GaAs quantum dots, hyperfine interaction with nuclear spins is the dominant source of spin decay for weak magnetic fields, with $T_{2}\simeq \mu{\rm s}$ and $T_{1}$ increasing from $\mu{\rm s}$ to ${\rm ms}$ with increasing magnetic field \cite{Elz04,Pet05}. For the transport problem in an open quantum dot considered here, the decoherence time $\tau_{\phi}$ of the whole wave function, rather than just its spin-dependent part, is the relevant quantity. Typically, $\tau_{\phi}$ is dominated by dephasing of the orbital degrees of freedom by electron-electron interactions.

Experiments \cite{Hui98} on the effect of a finite $\tau_{\phi}$ on spin-independent conduction have been analyzed in the past using B\"{u}ttiker's voltage probe model \cite{But88,Bar95,Bro97}. Extensions to spin-dependent  conduction have been proposed more recently \cite{Mic05,San06}. As described in Ref.\ \cite{Mic05}, one needs two types of voltage probes to describe spin relaxation and decoherence. One type of voltage probe is connected to a normal metal reservoir, while the other type of voltage probe is connected to a pair of ferromagnetic reservoirs (of opposite polarization, parallel to the polarization of the spin filters in the quantum point contacts). For each reservoir, an electron that enters it is reinjected into the quantum dot with a random phase. The ferromagnetic reservoirs conserve the spin (contributing only to $\tau_{\phi}$), while the normal metal reservoir randomizes the spin (contributing both to $T_{1}$ and $\tau_{\phi}$).

Each voltage probe is connected to the quantum dot by a tunnel barrier. The normal metal voltage probe has $N_{n}^{\uparrow}=N_{n}^{\downarrow}\equiv N_{n}$ channels for each spin direction and the ferromagnetic voltage probes have $N_{f}^{\uparrow}=N_{f}^{\downarrow}\equiv N_{f}$ channels. Each barrier has tunnel probability $\Gamma$ per channel and per spin direction. By taking the limit $\Gamma\rightarrow 0$, $N_{n},N_{f}\rightarrow\infty$ at fixed (dimensionless) tunnel conductances $\gamma_{n} = N_{n}\Gamma$, $\gamma_{f} = N_{f}\Gamma$ we ensure that the decay processes are spatially homogeneous \cite{Bro97}. The decay times are
\begin{equation}
T_{1}=\frac{h}{\gamma_{n} \Delta},\;\;
\tau_{\phi}=\frac{h}{(\gamma_{n}+\gamma_{f}) \Delta}\equiv\frac{h}{\gamma_{\phi}\Delta},
\label{T1T2def}
\end{equation}
with $\Delta$ the mean spacing of spin-degenerate levels and $\gamma_{\phi}\equiv\gamma_{n}+\gamma_{f}$. These time scales should be compared with the spin-dependent mean dwell time $\tau_{\rm dwell}^{\sigma}$ in the quantum dot without voltage probes, given by $\tau_{\rm dwell}^{\uparrow}=h/2\Delta$, $\tau_{\rm dwell}^{\downarrow}=h/\Delta$.

The electron reservoirs connected to the quantum dot have electrochemical potentials $\mu_{X}$, with $X=s$ (source), $X=d$ (drain), $X=n$ (normal metal voltage probe), and $X=f$ (ferromagnetic voltage probe). In the latter case we distinguish the two spin polarizations by a superscript: $\mu_{f}^{\uparrow},\mu_{f}^{\downarrow}$. We choose the zero of energy such that $\mu_{d}=0$, hence $\mu_{s}=eV$. Both the temperature and the applied voltage $V$ are assumed to be small compared to $\Delta$, so that we may neglect the energy dependence of the scattering processes.

The potentials of the voltage probes are determined by demanding that no current is drawn from the quantum dot \cite{But88},
\begin{eqnarray}
0&=&(2N_{n}-T_{n\rightarrow n}^{\uparrow}-T_{n\rightarrow n}^{\downarrow})\mu_{n}-T_{s\rightarrow n}^{\uparrow}eV-T_{f\rightarrow n}^{\uparrow}\mu_{f}^{\uparrow}\nonumber\\
&&\mbox{}-T_{f\rightarrow n}^{\downarrow}\mu_{f}^{\downarrow},\label{muneq}\\
0&=&(N_{f}-T_{f\rightarrow f}^{\uparrow})\mu_{f}^{\uparrow}-T_{s\rightarrow f}^{\uparrow}eV-T_{n\rightarrow f}^{\uparrow}\mu_{n},\label{mufupeq}\\
0&=&(N_{f}-T_{f\rightarrow f}^{\downarrow})\mu_{f}^{\downarrow}-T_{n\rightarrow f}^{\downarrow}\mu_{n}.\label{mufdowneq}
\end{eqnarray}
The current $I$ through the quantum dot then follows from
\begin{equation}
\frac{h}{e}I=(1-T_{s\rightarrow s}^{\uparrow})eV-T_{n\rightarrow s}^{\uparrow}\mu_{n}-T_{f\rightarrow s}^{\uparrow}\mu_{f}^{\uparrow}.\label{Ieq}
\end{equation}
Here $T_{X\rightarrow Y}^{\uparrow}$ and $T_{X\rightarrow Y}^{\downarrow}$ denote the transmission probabilities, summed over all channels, from reservoir $X$ to reservoir $Y$ with spin up or down. They satisfy the sum rules \cite{But88}
\begin{equation}
\sum_{Y=s,d,n,f}T^{\sigma}_{X\rightarrow Y}=\sum_{Y=s,d,n,f}T^{\sigma}_{Y\rightarrow X}=N_{X},\label{sumrules}
\end{equation}
with $N_{s}=N_{d}\equiv 1$. For later use we define
\begin{equation}
{\cal R}^{\uparrow}=2-T_{s\rightarrow s}^{\uparrow}-T_{d\rightarrow d}^{\uparrow}-T_{s\rightarrow d}^{\uparrow}-T_{d\rightarrow s}^{\uparrow},\;\;
{\cal R}^{\downarrow}=1-T_{d\rightarrow d}^{\downarrow}.\label{calRdef}
\end{equation}

Because of the spatial homogeneity of the coupling of the quantum dot to the voltage probes, the transmission probabilities for normal and ferromagnetic probes are related by ratios of tunnel conductances,
\begin{eqnarray}
&&\frac{T_{X\rightarrow n}^{\sigma}}{T_{X\rightarrow f}^{\sigma}}=\frac{T_{n\rightarrow X}^{\sigma}}{T_{f\rightarrow X}^{\sigma}}=\frac{\gamma_{n}}{\gamma_{f}},\;\;{\rm if}\;\;
X\in\{s,d\},\label{Tnfrelation1}\\
&&\frac{T_{X\rightarrow Y}^{\sigma}-\delta_{XY}N_{X}(1-\Gamma^{\sigma}_{\rm eff})}{T_{X'\rightarrow Y'}^{\sigma}-\delta_{X'Y'}N_{X'}(1-\Gamma^{\sigma}_{\rm eff})}=\frac{\gamma_{X}\gamma_{Y}}{\gamma_{X'}\gamma_{Y'}},\nonumber\\
&&\quad\quad{\rm if}\;\;X,Y\in\{n,f\}.\label{Tnfrelation2}
\end{eqnarray}
The effective tunnel probability $\Gamma^{\sigma}_{\rm eff}=(\Gamma\Delta)\rho^{\sigma}$ differs from the bare tunnel probability $\Gamma$ because the density of states $\rho^{\sigma}$ in the quantum dot has spin and energy dependent fluctuations around the average $1/\Delta$.

With the help of these relations the solution of Eqs.\ (\ref{muneq}--\ref{Ieq}) for the conductance $G=I/V$ can be written in terms of transmission probabilities between source and drain,
\begin{widetext}
\begin{eqnarray}
G&=&\frac{e^{2}}{h}\bigl[1-T_{s\rightarrow s}^{\uparrow}-{\cal Q}(1-T_{s\rightarrow s}^{\uparrow}-T_{d\rightarrow s}^{\uparrow})(1-T_{s\rightarrow s}^{\uparrow}-T_{s\rightarrow d}^{\uparrow})\bigr],\label{GQresult}\\
{\cal Q}&=&\frac{\Gamma^{\uparrow}_{\rm eff}\Gamma^{\downarrow}_{\rm eff}\gamma_{n}(\gamma_{n}+\gamma_{f})+\Gamma(\Gamma^{\uparrow}_{\rm eff}+\Gamma^{\downarrow}_{\rm eff})\gamma_{f}{\cal R}^{\downarrow}}{\Gamma^{\uparrow}_{\rm eff}\Gamma^{\downarrow}_{\rm eff}\gamma_{n}(\gamma_{n}+\gamma_{f})({\cal R}^{\uparrow}+{\cal R}^{\downarrow})+\Gamma(\Gamma^{\uparrow}_{\rm eff}+\Gamma^{\downarrow}_{\rm eff})\gamma_{f}{\cal R}^{\uparrow}{\cal R}^{\downarrow}}.\label{Qdef}
\end{eqnarray}
\end{widetext}

The transmission probabilities between source and drain are constructed from two scattering matrices $S^{\uparrow}$ and $S^{\downarrow}$, one for spin-up and one for spin-down. The spin-up scattering matrix is a $2\times 2$ matrix,
\begin{equation}
S^{\uparrow}=\begin{pmatrix}
r&t'\\
t&r'
\end{pmatrix},\label{Supdef}
\end{equation}
such that $T_{s\rightarrow s}^{\uparrow}=|r|^{2}$, $T_{d\rightarrow d}^{\uparrow}=|r'|^{2}$, $T_{s\rightarrow d}^{\uparrow}=|t|^{2}$, and $T_{d\rightarrow s}^{\uparrow}=|t'|^{2}$. The matrix $S$ is symmetric for $\beta=1$, meaning that $T_{s\rightarrow d}=T_{d\rightarrow s}$ in that case. For $\beta=2$ the two transmission probabilities are not related. Because of the voltage probes, $S$ is sub-unitary. The eigenvalues $\tau_{1},\tau_{2}\in[0,1]$ of the matrix $\openone-S^{\uparrow}S^{\uparrow\dagger}$ give the probability to enter one of the voltage probes.

The statistics of the matrix $S^{\uparrow}$ in an ensemble of chaotic quantum dots was calculated in Ref.\ \cite{Bro97} using the methods of random-matrix theory. It is given in terms of the polar decomposition
\begin{equation}
S^{\uparrow}=u\begin{pmatrix}
\sqrt{1-\tau_{1}}&0\\
0&\sqrt{1-\tau_{2}}
\end{pmatrix}u',\label{spolar}
\end{equation}
with unitary matrices $u'=u^{T}$ if $\beta=1$ and $u'$ independent of $u$ if $\beta=2$. These matrices are uniformly distributed in the unitary group. The distribution $P_{\beta}(\tau_{1},\tau_{2})$ is the Laguerre ensemble for $\gamma_{\phi}\ll 1$ and a more complicated (but known) function for larger $\gamma_{\phi}$.

In addition to $S^{\uparrow}$ we also need $S^{\downarrow}$. This is a single complex number, such that $T_{d\rightarrow d}^{\downarrow}=|S^{\downarrow}|^{2}$. It is constructed from the coefficients $r,t,t'$ in Eq.\ (\ref{spolar}) by reflecting spin-down from the source contact,
\begin{equation}
S^{\downarrow}=r'+\frac{e^{i\alpha}tt'}{1-e^{i\alpha}r}.\label{Sdowndef}
\end{equation}
(The phase shift $\alpha$ need not be specified because it drops out upon averaging over $u$ and $u'$.) Using Eq.\ (\ref{Sdowndef}) the statistics of $S^{\downarrow}$ follows from the statistics of $S^{\uparrow}$.

To complete the random-matrix theory, we need to know the statistics of the density of states $\rho^{\sigma}$ of the open quantum dot, which determines the effective tunnel probabilities. For weak decoherence we have the relation \cite{Bee01}
\begin{equation}
{\rm Tr}\,(\openone-S^{\sigma}S^{\sigma\dagger})=\rho^{\sigma}\gamma_{\phi}\Delta+{\cal O}(\gamma_{\phi}^{2}).\label{Rrhorelation}
\end{equation}
Since the left-hand-side of Eq.\ (\ref{Rrhorelation}) equals ${\cal R}^{\sigma}$ by definition (\ref{calRdef}), we have
\begin{equation}
\Gamma_{\rm eff}^{\sigma}/\Gamma\equiv\rho^{\sigma}\Delta={\cal R}^{\sigma}/\gamma_{\phi},\;\;{\rm if}\;\;\gamma_{\phi}\ll 1.\label{Gammaeffsmallgamma}
\end{equation}
In the opposite limit $\gamma_{\phi}\gg 1$ the fluctuations in the density of states can be neglected, so that
\begin{equation}
\Gamma_{\rm eff}^{\sigma}/\Gamma\equiv\rho^{\sigma}\Delta=1,\;\;{\rm if}\;\;\gamma_{\phi}\gg 1.\label{Gammaefflargegamma}
\end{equation}
These two limits are sufficient for the purpose of comparing coherent and incoherent regimes.

We calculate the average conductance $\langle G\rangle$ separately in the regime $\gamma_{f}\gg 1$ of strong orbital dephasing and the regime $\gamma_{f}\ll 1$ of weak orbital dephasing. For strong dephasing we have $\Gamma_{\rm eff}^{\sigma}\rightarrow\Gamma$, ${\cal R}^{\uparrow}\rightarrow 2$, ${\cal R}^{\downarrow}\rightarrow 1$, $\tau_{1},\tau_{2}\rightarrow 1$, hence
\begin{equation}
\langle G\rangle=\frac{2e^{2}}{h}\,\frac{1+\gamma_{n}}{4+3\gamma_{n}},\;\;{\rm if}\;\;\gamma_{f}\gg 1.\label{deltainc}
\end{equation}
This incoherent regime is insensitive to the presence or absence of time-reversal symmetry. By writing Eq.\ (\ref{deltainc}) as $\langle G\rangle=(e^{2}/h)[2(1-p)+\frac{3}{2}p]^{-1}$ with $p=\gamma_{n}/(1+\gamma_{n})$, we can understand it as a classical series resistance, weighted by the probability $p$ of a spin-flip scattering event. 

Turning now to the phase coherent regime, we find to linear order in $\gamma_{f}$ and $\gamma_{n}$ the expansions
\begin{equation}
\frac{h}{e^{2}}\langle G\rangle=\left\{\begin{array}{cl}
\tfrac{1}{3}+0.14\,\gamma_{n}+\tfrac{1}{24}\gamma_{f},&{\rm if}\;\;\beta=1,\\
\tfrac{1}{2}+0.10\,\gamma_{n},&{\rm if}\;\;\beta=2.
\end{array}\right.\label{Gweak}
\end{equation}
Note the absence of a term linear in $\gamma_{f}$ for $\beta=2$. The difference between the zeroth order terms $1/3$ and $1/2$ in the presence and absence of time-reversal symmetry is known as weak localization or coherent backscattering.

\begin{figure}
\centerline{\includegraphics[width=8cm]{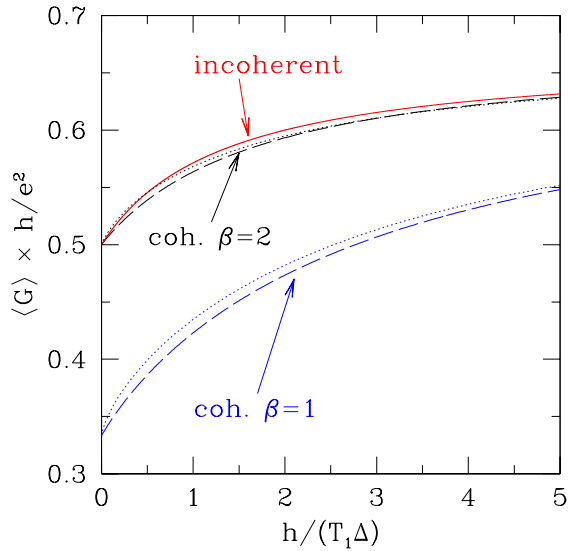}}
\caption{
(color online) Dependence of the average conductance $\langle G\rangle$ on the spin relaxation time $T_{1}$, normalized by the mean level spacing $\Delta$. The solid curve (red) is the incoherent result (\ref{deltainc}), valid for strong orbital dephasing. The dashed and dotted curves are the results of random-matrix theory for weak orbital dephasing, in the two cases of broken ($\beta=2$, black) and unbroken ($\beta=1$, blue) time reversal symmetry. The dashed curves are grand-canonical averages ($e^{2}/C\ll\Delta$) and the dotted curves are canonical averages ($e^{2}/C\gg\Delta$). The black and red curves lie close together, demonstrating that $T_{1}$ can be determined accurately from $\langle G\rangle$ for $\beta=2$. The large difference between the blue and red curves prevents this for $\beta=1$.
}
\label{fig_Gplot}
\end{figure}

In the absence of any orbital dephasing, $\gamma_{f}=0$, we obtain the results plotted in Fig.\ \ref{fig_Gplot} (dashed curves). Comparison with the incoherent result (\ref{deltainc}) (solid curve) shows that the presence or absence of orbital dephasing does not change $\langle G\rangle$ by more than a few \% if $\beta=2$ (no time-reversal symmetry). For $\beta=1$, in contrast, the dependence on $T_{1}$ is entirely different with and without orbital dephasing.

So far we have not included the effects of a finite charging energy $e^{2}/C$ (with $C$ the capacitance of the quantum dot). These results therefore apply to the regime $e^{2}/C\ll\Delta$. In the opposite, more realistic, regime $e^{2}/C\gg\Delta$ the charging energy introduces a weight factor equal to the density of states in the ensemble averages \cite{Bro97b}. This weight factor converts the grand-canonical average $\langle\cdots\rangle$ (considered so far) into a canonical average:
\begin{equation}
\langle\cdots\rangle_{\rm canonical}=\tfrac{1}{2}\Delta\langle\cdots (\rho^{\uparrow}+\rho^{\downarrow})\rangle.\label{canonical}
\end{equation}
The incoherent result (\ref{deltainc}) is the same in the canonical and grand-canonical ensembles, because the fluctuations in the density of states are suppressed by decoherence. In order to assess the importance of density-of-states fluctuations in the coherent regime, we approximate $(\Delta/2)(\rho^{\uparrow}+\rho^{\downarrow})\simeq ({\cal R}^{\uparrow}+{\cal R}^{\downarrow})/\langle{\cal R}^{\uparrow}+{\cal R}^{\downarrow}\rangle$. This formula interpolates smoothly between the two exact limits (\ref{Gammaeffsmallgamma}) and (\ref{Gammaefflargegamma}) of weak and strong decoherence. As shown in Fig.\ \ref{fig_Gplot} (dotted curves), the effect on the average conductance remains relatively small.

In conclusion, we have calculated the dependence on the spin relaxation time $T_{1}$ of the average conductance $\langle G\rangle$ of a quantum dot with a spin-filtering quantum point contact. In the incoherent regime there is a simple one-to-one relationship (\ref{deltainc}) between $\langle G\rangle$ and $T_{1}$. The presence or absence of orbital dephasing was found to be insignificant for $\beta=2$, so that the value of $T_{1}$ can be extracted from $\langle G\rangle$ with good accuracy --- without requiring knowledge of coherence time or charging energy. For $\beta=1$, in contrast, the interplay with the weak localization effect obscures the effect of spin relaxation.

I am indebted to C. M. Marcus for drawing my attention to Ref.\ \cite{Zum01} and for valuable comments on the manuscript. This research was supported by the Dutch Science Foundation NWO/FOM.


\begin{thebibliography}{99}
\bibitem{Zai05} O. Zaitsev, D. Frustaglia, and K. Richter, Phys.\ Rev.\ Lett.\ {\bf 94}, 026809 (2005).
\bibitem{Sau04} O. Sauret and D. Feinberg, Phys.\ Rev.\ Lett.\ {\bf 92}, 106601 (2004).
\bibitem{Zum01} D. M. Zumb\"{u}hl, J. A. Folk, J. B. Miller, S. K. Watson, C. M. Marcus, S. R. Patel, C. I. Duru\"{o}z, and J. S. Harris, Jr., 2001 March Meeting Bulletin of the American Physical Society, Abstract C25.006.
\bibitem{Sac01}  A. S. Sachrajda, P. Hawrylak, M. Ciorga, C. Gould, and P. Zawadzki, Physica E {\bf 10}, 493 (2001).
\bibitem{Pot02} R. M. Potok, J. A. Folk, C. M. Marcus, and V. Umansky, Phys.\ Rev.\ Lett.\ {\bf 89}, 266602 (2002).
\bibitem{Fal02} V. I. Fal'ko and T. Jungwirth, Phys. Rev. B {\bf 65}, 081306 (2002); D. M. Zumb\"{u}hl, et al., Phys.\ Rev.\ B {\bf 69}, 121305(R) (2004).
\bibitem{Eng04} H.-A. Engel, L. P. Kouwenhoven, D. Loss, and C. M. Marcus, 
Quantum Inf.\ Process.\ {\bf 3}, 115 (2004).
\bibitem{Gol04} V. N. Golovach, A. Khaetskii, and D. Loss, Phys.\ Rev.\ Lett.\ {\bf 93}, 016601 (2004).
\bibitem{Coi05} W. A. Coish and D. Loss, Phys.\ Rev.\ B {\bf 72}, 125337 (2005).
\bibitem{Elz04} J. M. Elzerman, R. Hanson, L. H. Willems van Beveren, B. Witkamp, L. M. K. Vandersypen, and L. P. Kouwenhoven, Nature {\bf 430}, 431 (2004).
\bibitem{Pet05} J. R. Petta, A. C. Johnson, J. M. Taylor, E. A. Laird, A. Yacoby, M. D. Lukin, C. M. Marcus, M. P. Hanson, and A. C. Gossard, Science {\bf 309}, 2180 (2005).
\bibitem{Hui98} A. G. Huibers, M. Switkes, C. M. Marcus, K. Campman, and A. C. Gossard, Phys.\ Rev.\ Lett.\ {\bf 81}, 200 (1998); A. G. Huibers, et al., Phys.\ Rev.\ Lett.\ {\bf 83}, 5090 (1999).
\bibitem{But88} M. B\"{u}ttiker, IBM J. Res.\ Dev.\ {\bf 32}, 63 (1988).
\bibitem{Bar95} H. U. Baranger and P. A. Mello, Phys. Rev. B {\bf 51}, 4703 (1995); P. W. Brouwer and C. W. J. Beenakker, Phys. Rev. B {\bf 51}, 7739 (1995).
\bibitem{Bro97} P. W. Brouwer and C. W. J. Beenakker, Phys.\ Rev.\ B {\bf 55}, 4695 (1997); {\bf 66}, 209901(E) (2002).
\bibitem{Mic05} B. Michaelis and C. W. J. Beenakker, cond-mat/0512465.
\bibitem{San06} P. San-Jose and E. Prada, cond-mat/0601365.
\bibitem{Bee01} C. W. J. Beenakker and P. W. Brouwer, Physica E {\bf 9}, 463 (2001).
\bibitem{Bro97b} P. W. Brouwer, S. A. van Langen, K. M. Frahm, M. B\"{u}ttiker, and C. W. J. Beenakker, Phys.\ Rev.\ Lett.\ {\bf 79}, 913 (1997).
\end{thebibliography}
\end{document}